\documentclass[twocolumn,floatfix,showpacs,eqsecnum,prb]{revtex4}
\usepackage{graphicx}
\usepackage{amsmath}
\usepackage{amssymb}
\usepackage{bm}

\begin{document}

\title{Nonzero temperature effects on antibunched photons emitted by a quantum point contact out of equilibrium}
\author{I. C. Fulga}
\affiliation{Instituut-Lorentz, Universiteit Leiden, P.O. Box 9506, 2300 RA Leiden, The Netherlands}
\author{F. Hassler}
\affiliation{Instituut-Lorentz, Universiteit Leiden, P.O. Box 9506, 2300 RA Leiden, The Netherlands}
\author{C. W. J. Beenakker}
\affiliation{Instituut-Lorentz, Universiteit Leiden, P.O. Box 9506, 2300 RA Leiden, The Netherlands}
\date{January 2010}
\begin{abstract}
Electrical current fluctuations in a single-channel quantum point contact can produce photons (at frequency $\omega$ close to the applied voltage $V\times e/\hbar$) which inherit the sub-Poissonian statistics of the electrons.  We extend the existing zero-temperature theory of the photostatistics to nonzero temperature $T$. The Fano factor ${\cal F}$ (the ratio of the variance and the average photocount) is $<1$ for $T<T_{c}$ (antibunched photons) and $>1$ for $T>T_{c}$ (bunched photons). The crossover temperature $T_{c}\simeq\Delta\omega\times\hbar/k_{B}$ is set by the band width $\Delta\omega$ of the detector, even if $\hbar\Delta\omega\ll eV$. This implies that narrow-band detection of photon antibunching is hindered by thermal fluctuations even in the low-temperature regime where thermal electron noise is negligible relative to shot noise.
\end{abstract}
\maketitle

\section{Introduction}
\label{intro}

In classical mechanics, time dependent fluctuations of an electrical current produce photons with the Poisson statistics of classical particles.\cite{Gla63} The variance ${\rm Var}\,n$ of the number $n$ of photons detected in a time $t_{\rm det}$ is then equal to the mean $\langle n\rangle$. Quantum mechanics changes the photostatistics.\cite{Man05} The bosonic nature of the photons would naturally lead to photon bunching, with ${\rm Var}\,n>\langle n\rangle$. Photon antibunching, with ${\rm Var}\,<\langle n\rangle$, is also possible, if the photons can inherit the sub-Poissonian statistics of the electrons.\cite{Bee01}

It is an experimental challenge to detect antibunched photons produced by electronic shot noise in a quantum conductor.\cite{Gab04,Zak10} The theoretical prediction\cite{Bee04} is that photons emitted by a single-channel quantum point contact should have a Fano factor ${\cal F}={\rm Var}\,n/\langle n\rangle$ smaller than unity at zero temperature, for frequencies $\omega$ close to the applied voltage $V\times e/\hbar$. More specifically,
\begin{equation}
{\cal F}=1-\tfrac{2}{3}(\gamma_{0}\Delta\omega)\tau(1-\tau)\label{Foldresult}
\end{equation}
for photodetection with efficiency $\gamma_{0}$ in the frequency interval $(eV/\hbar-\Delta\omega,eV/\hbar)$. The transmission probability $\tau$ through the quantum point contact is assumed to be energy independent on the scale of $eV$. Eq.\ \eqref{Foldresult} is derived in the limit of weak coupling ($\gamma_{0}\Delta\omega\ll 1$) of electrons to photons, so that the deviations from Poisson statistics remain small. It is also assumed that the photons can be detected individually, see Ref.\ \onlinecite{Leb09} for an alternative detection scheme.

It is the purpose of the present paper to extend the theory of Ref.\ \onlinecite{Bee04} to nonzero temperatures, in order to identify the conditions on the temperature needed to observe the photon antibunching. Clearly, photon bunching should take over when the electrical shot noise drops below the thermal noise, which happens when $k_{B}T$ becomes larger than $eV$. While $k_{B}T<eV$ is the condition for photon antibunching in the case of wide-band detection, a more stringent condition $k_{B}T<\hbar\Delta\omega$ holds for narrow-band detection. 

More precisely, we obtain a crossover temperature $T_{c}\approx\hbar\Delta\omega/4k_{B}$ at which ${\cal F}=1$ for $\Delta\omega\ll eV$. In this low-temperature regime shot noise still dominates over thermal noise, yet the photon antibunching is lost. One qualitative way to understand this is, is to compare the coherence time $t_{\rm coh}\simeq 1/\Delta\omega$ of the detected radiation with the coherence time $t_{T}\simeq\hbar/k_{B}T$ of thermally excited electron-hole pairs. For $t_{\rm coh}>t_{T}$ the detected photons result from many uncorrelated electron-hole recombination events, and the one-to-one relationship between electron and photon statistics is lost.

In the next section we give the nonzero temperature generalization of the theory of Ref.\ \onlinecite{Bee04}, and then in Sec.\ \ref{shotnoise} we specialize to the shot-noise regime $k_{B}T\ll eV$. General results in both the shot noise and thermal noise regimes are presented in Sec.\ \ref{general}. Technical details are summarized in the Appendices.

\section{Generating function at nonzero temperature}
\label{nonzero}

We seek the nonzero-temperature generalization of the formula\cite{Bee04} 
\begin{equation}
F(\xi)=\biggl\langle\prod_{m=1}^{N}{\rm Det}\biggl(1+T_{m}[e^{Z}e^{Z^{\dagger}}-1]\biggr)\biggr\rangle,\label{FxiresultzeroT}
\end{equation}
for the factorial-moment generating function $F(\xi)$ of the photocount. We first introduce the notation and then present the required generalization.

The photons are produced by time-dependent current fluctuations in a quantum point contact, characterized by transmission eigenvalues $T_{1},T_{2},\ldots T_{N}$, with $N$ the number of propagating electronic modes. The current flows between two reservoirs, with Fermi functions
\begin{align}
&f_{L}(\varepsilon)=\bigl(1+\exp[(\varepsilon-eV-E_{F})/k_{B}T]\bigr)^{-1},\label{fLdef}\\
&f_{R}(\varepsilon)=\bigl(1+\exp[(\varepsilon-E_{F})/k_{B}T]\bigr)^{-1}.\label{fRdef}
\end{align}
The current fluctuations can be due to thermal noise (at temperature $T$) or due to shot noise (at a voltage $V$ applied over the point contact). We take the transmission eigenvalues $T_{m}$ as energy independent in the range $\max(eV,k_{B}T)$ near the Fermi energy $E_{F}$.

The photons are detected during a time $t_{\rm det}$ in a narrow frequency interval $\Delta\omega$ around frequency $\Omega$, as determined by the detection efficiency $\gamma(\omega)$. Antibunching of the photons requires that $\Omega$ is tuned to the applied voltage, $\Omega\simeq eV/\hbar$. (In the following we set $\hbar$ and $e$ equal to unity.)

The average $\langle\cdots\rangle$ in Eq.\ \eqref{FxiresultzeroT} indicates a Gaussian integration over the complex numbers $z_{p}$,
\begin{equation}
\langle\cdots\rangle=\prod_{p}\frac{\gamma_{p}}{\pi}\int d^{2}z_{p}e^{-\gamma_{p}|z_{p}|^{2}}\ldots.\label{Gaussiandef}
\end{equation}
The matrix $Z$  has elements $Z_{pp'}=\xi^{1/2}z_{p-p'}\gamma_{p-p'}$, depending only on the difference of the indices $p$ and $p'$. This difference represents the discretized frequency $\omega_{p-p'}=(p-p')\times 2\pi/t_{\rm det}$ of a photon emitted by an electronic transition from energy $\varepsilon_{p}$ to $\varepsilon_{p'}$ and detected with efficiency $\gamma_{p-p'}=(2\pi/t_{\rm det})\gamma(\omega_{p-p'})$. Since $\gamma(\omega)\equiv 0$ for $\omega\leq 0$, the matrix $Z$ is a lower-triangular matrix. The discretization of frequency and energy is eliminated at the end of the calculation, by taking the limit $t_{\rm det}\rightarrow\infty$.

The expansion
\begin{equation}
F(\xi)=\sum_{k=0}^{\infty}\frac{\xi^{k}}{k!}\langle n^{k}\rangle_{f}\label{Fxiexpansion}
\end{equation}
of $F(\xi)$ in powers of $\xi$ gives the factorial moments $\langle n^{k}\rangle_{f}=\langle n(n-1)(n-2)\cdots(n-k+1)\rangle$ of the number of detected photons. Antibunching means that the variance of the photocount ${\rm Var}\,n=\langle n^{2}\rangle-\langle n\rangle^{2}$ is smaller than the average, or equivalently that the Fano factor ${\cal F}={\rm Var}\,n/\langle n\rangle<1$. 

As outlined in App.\ \ref{outline}, at nonzero temperature we have instead of Eq.\ \eqref{FxiresultzeroT} the generating function
\begin{widetext}
\begin{align}
F(\xi)&=\left\langle\prod_{m=1}^{N}{\rm Det}\begin{pmatrix}
1+T_{m} f_{L}(e^{Z}e^{Z^{\dagger}}-1)&\sqrt{T_{m} (1-T_{m} )}f_{L}(e^{-Z^{\dagger}}-e^{Z})\\
\sqrt{T_{m} (1-T_{m} )}f_{R}(e^{-Z}-e^{Z^{\dagger}})&1+T_{m} f_{R}(e^{-Z}e^{-Z^{\dagger}}-1)
\end{pmatrix}
\right\rangle.\label{Fxi1}
\end{align}
\end{widetext}
The Fermi function $f_{L}(\varepsilon)$ in the left electronic reservoir is contained in the diagonal matrix $f_{L}$, with elements $(f_{L})_{pp'}=\delta_{pp'}f_{L}(\varepsilon_{p})$, $\varepsilon_{p}=p\times 2\pi/t_{\rm det}$. Similarly, the Fermi function $f_{R}(\varepsilon)$ in the right reservoir is contained in the diagonal matrix $f_{R}$.

Following the steps in App.\ \ref{outline}, the expression \eqref{Fxi1} can be reduced to the more compact form
\begin{equation}
F(\xi)=\biggl\langle\prod_{m=1}^{N}{\rm Det}\biggl(1+T_{m}[\bar{f}_{R}e^{Z^{\dagger}}f_{L}-f_{R}e^{-Z}\bar{f_{L}}]{\cal M}\biggr)\biggr\rangle,\label{Fxiresult}
\end{equation}
with the definitions $\bar{f}_{L}=1-f_{L}$, $\bar{f}_{R}=1-f_{R}$, ${\cal M}=e^{Z}-e^{-Z^{\dagger}}$. The zero-temperature limit \eqref{FxiresultzeroT} follows from Eq.\ \eqref{Fxiresult} by setting $f_{L}=1$, $f_{R}=0$ in the energy interval $E_{F}<\varepsilon<E_{F}+V$. (There are no current fluctuations outside of this energy interval for $T=0$.)

\section{Shot noise regime}
\label{shotnoise}

The result \eqref{Fxiresult} holds for any temperature, provided that the energy dependence of the transmission eigenvalues may be neglected. In particular, it describes both thermal noise and shot noise. A simpler formula is obtained in the shot noise regime $k_{B}T\ll V$. Thermal noise can then be neglected and only the finite temperature effects on the shot noise are retained. We assume $\Delta\omega\ll\Omega\simeq V$, so even if $k_{B}T\ll V$, the relative magnitude of $\Delta\omega$ and $k_{B}T$ is still arbitrary.

\subsection{Generating function}
\label{Fshotnoise}

The first simplification in this regime is that we may set $f_{R}e^{-Z}\bar{f}_{L}\rightarrow 0$, since $f_{R}(\varepsilon)\bar{f}_{L}(\varepsilon')\rightarrow 0$ for $\varepsilon'\leq\varepsilon$. Eq.\ \eqref{Fxiresult} reduces to
\begin{equation}
F(\xi)=\biggl\langle\prod_{m=1}^{N}{\rm Det}\biggl(e^{-Z^{\dagger}}+T_{m}f_{L}{\cal M}\bar{f}_{R}\biggr)\biggr\rangle,\label{Fxiresult1}
\end{equation}
where we have multiplied by ${\rm Det}\,e^{-Z^{\dagger}}=1$.

The second simplification is that we can ignore energies separated by $pV$ with $p\geq 2$, because $V$ is the largest energy scale in the problem. Since $Z^{p}$ and ${Z^{\dagger}}^{p}$ connect energies separated by $p\Omega\simeq pV$, we may set $Z^{p},{Z^{\dagger}}^{p}\rightarrow 0$ for $p\geq 2$. From Eq.\ \eqref{Fxiresult1} we arrive at
\begin{equation}
F(\xi)=\biggl\langle\prod_{m=1}^{N}{\rm Det}\biggl(1-Z^{\dagger}+T_{m}f_{L}(Z+Z^{\dagger})\bar{f}_{R}\biggr)\biggr\rangle.\label{Fxiresult2}
\end{equation}
Following the steps in App.\ \ref{outline2}, the determinant may be rewritten in the more convenient form (bilinear in $Z,Z^{\dagger}$),
\begin{equation}
F(\xi)=\biggl\langle\prod_{m=1}^{N}{\rm Det}\biggl(1+T_{m}(1-T_{m})f_{L}Z\bar{f}_{R}Z^{\dagger}\biggr)\biggr\rangle.\label{Fxiresult3}
\end{equation}

\subsection{Moment expansion}
\label{momentsshotnoise}

The generating function \eqref{Fxiresult3} is of the form $F(\xi)=\prod_{m}{\rm Det}\,(1+X_{m})$ with $X_{m}$ of order $\xi$. An expansion in powers of $\xi$ can be obtained by starting from the identity
\begin{equation}
{\textstyle\prod_{m}}{\rm Det} \,(1+X_m) = \exp\bigl[\textstyle{\sum_{m}}{\rm Tr}\,\ln(1+X_m)\bigr],\label{DetLogrelation}
\end{equation}
and expanding in turn, the logarithm and the exponential. Up to second order in $\xi$ we have the expansion
\begin{align}
F(\xi)={}& 1 + \bigl\langle {\textstyle\sum_{m}}{\rm Tr}\,X_{m}\bigr\rangle- \tfrac{1}{2}\bigl\langle {\textstyle\sum_{m}}{\rm Tr}\,X_{m}^{2}\bigr\rangle \nonumber \\ 
& + \tfrac{1}{2}\bigl\langle\bigl({\textstyle\sum_{m}}{\rm Tr}\,X_{m}\bigr)^{2} \bigr\rangle+{\cal O}(\xi^{3}),\label{Fxiexpansion2}
\end{align}
from which we can extract the first two factorial moments,
\begin{equation}
F(\xi)=1+\xi\langle n\rangle+\tfrac{1}{2}\xi^{2}\bigl(\langle n^{2}\rangle-\langle n\rangle\bigr)+{\cal O}(\xi^{3}).
\end{equation}

We perform the Gaussian averages and obtain the average photocount $\langle n\rangle$ and the variance ${\rm Var}\,n=\langle n^{2}\rangle-\langle n\rangle^{2}$ in the shot noise regime,
\begin{align}
\langle n\rangle={}&\frac{t_{\rm det}}{2\pi}S_{1}\int d\omega\,\gamma(\omega)\int d\varepsilon\, f_{L}(\varepsilon+\omega)\bar{f}_{R}(\varepsilon),\label{barnresult}\\
{\rm Var}\, n={}&\langle n \rangle + \frac{t_{\rm det}}{2\pi}S_{1}^{2}\int d\omega\,\biggl[\gamma(\omega)\int d\varepsilon\, f_{L}(\varepsilon+\omega)\bar{f}_{R}(\varepsilon)\biggr]^{2}\nonumber\\
&-\frac{t_{\rm det}}{2\pi}S_{2}\int d\varepsilon\,\biggl[f_{L}(\varepsilon)\int d\omega\,\gamma(\omega)\bar{f}_{R}(\varepsilon-\omega)\biggr]^{2}\nonumber\\
&-\frac{t_{\rm det}}{2\pi}S_{2}\int d\varepsilon\,\biggl[\bar{f}_{R}(\varepsilon)\int d\omega\,\gamma(\omega)f_{L}(\varepsilon+\omega)\biggr]^{2}.\label{secondcumulant}
\end{align}
We have defined
\begin{equation}
S_{p} = \sum_{m} [ T_{m}(1-T_{m})]^{p}.\label{Spdef}
\end{equation}

Since the two reservoirs are at the same temperature, we can write $f_{L}(\varepsilon)=f(\varepsilon-V-E_{F})$ and $\bar{f}_{R}=f(E_{F}-\varepsilon)$ in terms of a single Fermi function
\begin{equation}
f(\varepsilon)=(1+e^{\varepsilon/k_{B}T})^{-1}.\label{fdef}
\end{equation}
We abbreviate $\Gamma(\varepsilon,\omega)=\gamma(\omega)f(\varepsilon)f(\omega-\varepsilon-V)$ and can then write Eqs.\ \eqref{barnresult} and \eqref{secondcumulant} in the compact form
\begin{align}
\langle n\rangle={}&\frac{t_{\rm det}}{2\pi}S_{1}\int d\omega\int d\varepsilon\,\Gamma(\varepsilon,\omega),\label{barnresultb}\\
{\rm Var}\, n={}&\langle n \rangle + \frac{t_{\rm det}}{2\pi}\int d\omega\int d\varepsilon\, \Gamma(\varepsilon,\omega)\nonumber\\
&\times\biggl[ S_{1}^{2}\int d\varepsilon'\,\Gamma(\varepsilon',\omega)-2S_{2}\int d\omega'\,\Gamma(\varepsilon,\omega')\biggr].\label{secondcumulantb}
\end{align}
The difference ${\rm Var}\,n-\langle n\rangle$ contains a positive term $\propto S_{1}^{2}$ and a negative term $\propto S_{2}$. The sign of this difference determines whether there is bunching or antibunching of the detected photons.

\subsection{Crossover from antibunching to bunching}
\label{crossovershotnoise}

To investigate the crossover from antibunching to bunching with increasing temperature, we take a block-shaped response function
\begin{equation}
\gamma(\omega) = \left\{ \begin{array}{ll}
\gamma_0 & \text{if} \quad V-\Delta\omega < \omega < V ,\\
0 & \text{otherwise}.
\end{array} \right.\label{boxshape}
\end{equation}
In the low-temperature regime $k_{B}T\ll\Delta\omega$ the function $\Gamma(\varepsilon,\omega)$ then has a block shape as well and we recover the results 
\begin{align}
&\langle n\rangle=\frac{t_{\rm det}\Delta\omega}{2\pi}\gamma_{0}\Delta\omega\frac{1}{2}S_{1},\label{barnlow}\\
&{\rm Var} \, n - \langle n \rangle=\frac{t_{\rm det}\Delta\omega}{2\pi}(\gamma_{0}\Delta\omega)^{2}\frac{1}{3} (S_{1}^{2} - 2S_{2})\label{varnlow} 
\end{align}
of Ref.\ \onlinecite{Bee04}. These correspond to a Fano factor
\begin{equation}
{\cal F}=1+\tfrac{2}{3}\gamma_{0}\Delta\omega(S_{1} - 2S_{2}/S_{1}).\label{Flow}
\end{equation}
For a single-channel conductor $S_{2}=S_{1}^{2}$, so there is antibunching (${\cal F}<1$) at low temperatures. 

At high temperatures $k_{B}T\gg\Delta\omega$, but still in the shot-noise regime $k_{B}T\ll V$, we may substitute $\Gamma(\varepsilon,\omega)\rightarrow -\gamma(\omega)k_{B}Tdf(\varepsilon)/d\varepsilon$ into Eqs.\ \eqref{barnresultb} and \eqref{secondcumulantb}, which gives
\begin{align}
&\langle n\rangle=\frac{t_{\rm det}\Delta\omega}{2\pi}\gamma_{0}k_{B}TS_{1},\label{barnhigh}\\
&{\rm Var} \, n - \langle n \rangle=\frac{t_{\rm det}\Delta\omega}{2\pi}(\gamma_{0}k_{B}T)^{2}S_{1}^{2}.\label{varnhigh} 
\end{align}
The Fano factor
\begin{equation}
{\cal F}=1+\gamma_{0}k_{B}TS_{1}\label{Fhigh}
\end{equation}
is now $>1$ --- hence there is photon bunching.

\begin{figure}[htb]
\includegraphics[width=0.9\linewidth]{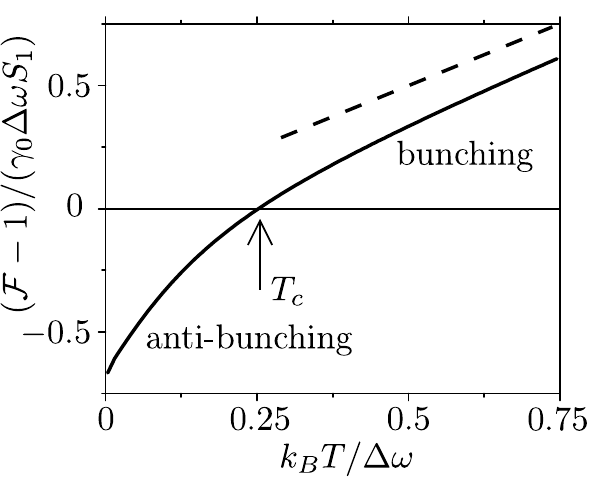}
\caption{Crossover with increasing temperature from antibunching (Fano factor ${\cal F}<1$) to bunching (${\cal F}>1$) of the photons produced by a single-channel quantum point contact in the shot noise regime ($k_{B}T\ll V$). The solid curve is calculated from Eqs.\  \eqref{barnresultb}--\eqref{boxshape}. The dashed line is the asymptote \eqref{Fhigh}. The crossover temperature $T_{c}$ from Eq.\ \eqref{kTc} is indicated.
\label{fig:limits}}
\end{figure}

The crossover temperature $T_{c}$, at which ${\cal F}=1$, can be calculated numerically from Eqs.\ \eqref{barnresultb} and \eqref{secondcumulantb}. In the single-channel case, when $S_{2}=S_{1}^{2}$, we find
\begin{equation}
k_{B}T_{c}\approx0.25\,\Delta\omega.\label{kTc}
\end{equation}
The crossover is shown graphically in Fig.\ \ref{fig:limits}.

\section{Beyond the shot noise regime}
\label{general}

\begin{figure}[htb]
\includegraphics[width=0.9\linewidth]{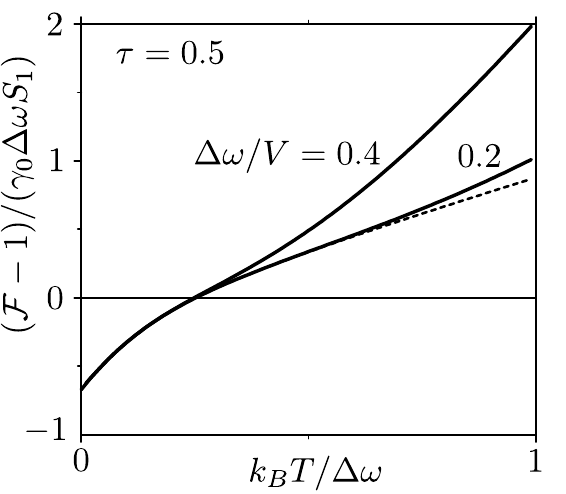}
\caption{Same as Fig.\ \ref{fig:limits}, but now without making the restriction to the shot noise regime (so without assuming $k_{B}T\ll V$). The two solid curves are calculated from Eq.\ \eqref{Fxiresult} for two values of $\Delta\omega/k_{B}T$ (both for the single-channel case with transmission probability $\tau=0.5$). Both curves converge to the shot noise result at low temperatures (shown dashed). \label{fig:comp}} 
\end{figure}

\begin{figure}[htb]
\includegraphics[width=0.9\linewidth]{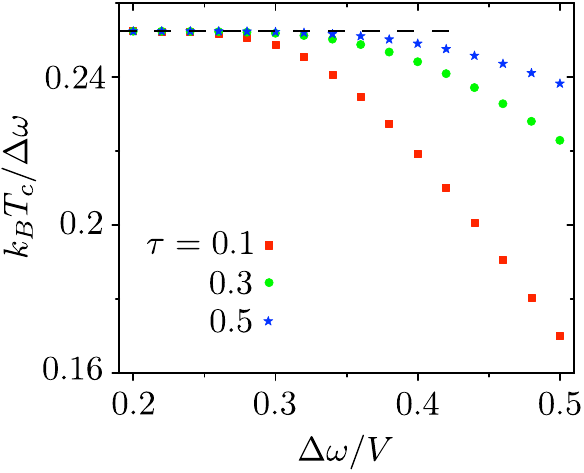}
\caption{Dependence of the crossover temperature $T_{c}$ (at which ${\cal F}=1$) on the band width $\Delta\omega$. The points are calculated from Eq.\ \eqref{Fxiresult} for three values of the single-channel transmission probability $\tau$. For $\Delta\omega\ll V$ all points converge to the shot noise limit \eqref{kTc} (dashed line). \label{fig:tc}} 
\end{figure}

In the previous section we assumed $k_{B}T\ll V$ (shot noise regime). For arbitrary relative magnitude of $k_{B}T$ and $V$, the general formula \eqref{Fxiresult} can be used. With the help of Eq.\ \eqref{DetLogrelation}, this general expression of the form ${\rm Det}\,(1+X)$ was expanded to second order in powers of $\xi$. In this case however, since $X={\cal O}(\sqrt{\xi})$, terms up to order $X^{4}$ had to be retained. The first two moments of $n$ are obtained as integrals over energy and frequency, similar to Eqs.\ \eqref{barnresultb} and \eqref{secondcumulantb} but containing many more terms in the integrands. The results shown in Figs.\ \ref{fig:comp} and \ref{fig:tc} are for the case $N=1$, $T_{1}=\tau$ of a single channel, and for the box-shaped response function \eqref{boxshape}.

As expected, all curves converge to the shot-noise results when $k_{B}T\ll V$ (shown dashed). At higher temperatures, the Fano factor lies above the shot noise limit due to the appearance of thermal noise. The temperature $T_{c}$ at which antibunching crosses over into bunching, so when ${\cal F}=1$, follows the shot noise limit \eqref{kTc} for narrow-band detection ($\Delta\omega\ll V$). With increasing band width, $T_{c}$ drops below the shot noise limit, in particular for small transmission probability $\tau$. For $\tau=0.5$ the shot noise limit remains accurate even for band widths $\Delta\omega$ as large as $V/2$.

\section{Conclusion}
\label{conclude}

In conclusion, we have investigated the effects of a nonzero temperature on the degree of antibunching of photons produced by current fluctuations in a quantum point contact. Antibunching crosses over into bunching as a result of thermal noise in the point contact, but this is not the dominant effect in the case of narrow-band detection. In that case the finite coherence time of electron-hole pairs governs the transition from photon antibunching to photon bunching, which occurs at a temperature $k_BT_{c}\simeq\Delta\omega$ even if $k_BT_{c}\ll V$ (so even if thermal noise is negligible relative to shot noise).

The optimal conditions for the observation of antibunched photons are reached for a band width $\Delta\omega\approx V/2$ and a transmission probability $\tau\approx 1/2$ through a single-channel quantum point contact. In that case $k_BT_{c}\approx V/8$ has the largest value at any given applied voltage.

\acknowledgments

We thank D. C. Glattli for a discussion which motivated this research. It was supported by the Dutch Science Foundation NWO/FOM and by the EU Network NanoCTM.

\appendix
\section{Derivation of the generating function at nonzero temperature}
\label{outline}

We briefly describe how the analysis of Ref.\ \onlinecite{Bee04} can be generalized to nonzero temperatures, in order to arrive at Eq.\ \eqref{Fxi1}. Referring to the equations in that paper, the first equation which changes is Eq.\ (5), which now reads
\begin{equation}
F(\xi) =\bigl\langle e^{-a^{\dagger}DZa}e^{b^{\dagger}DZb}e^{b^{\dagger}DZ^{\dagger}b}e^{-a^{\dagger}DZ^{\dagger}a}\bigr\rangle.\label{Fxi0}
\end{equation}
The four factors correspond to the four current operators that need to be taken into account: $I_\text{in}^\dagger$, $I_\text{out}^\dagger$, $I_\text{out}$, and $I_\text{in}$.

The operator $a^{\dagger}$ creates an incoming electron, while $b^{\dagger}$ creates an outgoing electron. The matrix $D$ projects on the right lead, where the current is evaluated. (Since $D$ commutes with $Z$, we can write $DZ$ instead of $DZD$.) One can relate $b=Sa$, with $S$ the unitary scattering matrix, so one can write the entire generating function in terms of the operators $a$. The expectation value $\langle\cdots\rangle$ is both an expectation value over the fermion operators $a$, as well as the average over the Gaussian variables $Z,Z^{\dagger}$.

Following the steps of Ref.\ \onlinecite{Bee04}, we calculate the expectation value of the fermion operators by means of the identity
\begin{align}
&\biggl\langle\prod_{n}e^{a^{\dagger}A_{n}a}\biggr\rangle={\rm Det}\,(1+AB),\label{identity2a}\\
&A=\Bigl(\prod_{n}e^{A_{n}}\Bigr)-1,\;\;B_{ij}=\langle a_{j}^{\dagger}a_{i}\rangle.\label{identity2b}
\end{align}
We have $B_{ij}=\delta_{ij}f_{i}$, with $f_{i}$ the Fermi occupation number in channel $i$. The matrix $A$ is given by $A=e^{X}e^{Y}e^{Y^{\dagger}} e^{X^{\dagger}} - 1$, with $X=-DZ$ and $Y=S^{\dagger}DZS$. Notice that $X^{p}=D(-Z)^{p}$ and $Y^{p}=S^{\dagger}DZ^{p}S$.

We now make the assumption of an energy independent scattering matrix, so $S,S^{\dagger}$ commute with $Z,Z^{\dagger}$. The determinant is invariant under a change of basis, and by working in the eigenchannel basis we can reduce $S$ to a $2\times 2$ matrix $S_{m}$ for each eigenchannel,
\begin{equation}
S_{m}=\begin{pmatrix}
\sqrt{1-T_{m}}&\sqrt{T_{m}}\\
\sqrt{T_{m}}&-\sqrt{1-T_{m}}
\end{pmatrix},\label{Smdef}
\end{equation}
with $T_{m}$, $m=1,2,\ldots N$ the transmission eigenvalue. The matrix structure of $f$, $D$, and $Z$ in this basis is
\begin{equation}
f=\begin{pmatrix}
f_{L}&0\\
0&f_{R}
\end{pmatrix},\;\;D=\begin{pmatrix}
0&0\\
0&1
\end{pmatrix},\;\;Z=\begin{pmatrix}
Z&0\\
0&Z
\end{pmatrix}.\label{fDdef}
\end{equation}
Substitution of Eqs.\ \eqref{identity2a}--\eqref{fDdef} into Eq.\ \eqref{Fxi0} leads after some algebraic manipulations to the result \eqref{Fxi1}.

The determinant in Eq.\ \eqref{Fxi1} can be reduced by means of the folding identity
\begin{equation}
{\rm Det}\,\begin{pmatrix}
M_{11}&M_{12}\\
M_{21}&M_{22}
\end{pmatrix}=
{\rm Det}M_{11}\,{\rm Det}\,(M_{22}-M_{21}M_{11}^{-1}M_{12}),\label{folding}
\end{equation}
leading to
\begin{widetext}
\begin{align}
F(\xi)&=\biggl\langle\prod_{m=1}^{N}{\rm Det}\bigl[1+T_{m}f_{L}(e^{Z}e^{Z^{\dagger}}-1)\bigr]
{\rm Det}\biggl(1+T_{m}f_{R}(e^{-Z}e^{-Z^{\dagger}}-1)\nonumber\\
&\qquad-T_{m}(1-T_{m})f_{R}(e^{-Z}-e^{Z^{\dagger}})[1+T_{m}f_{L}(e^{Z}e^{Z^{\dagger}}-1)]^{-1}f_{L}(e^{-Z^{\dagger}}-e^{Z})\biggr)\biggr\rangle.\label{Fxi2}
\end{align}

We continue the reduction of the determinant, using first the identity
\begin{equation}
[1+T_{m}f_{L}(e^{Z}e^{Z^{\dagger}}-1)]^{-1}f_{L}(e^{-Z^{\dagger}}-e^{Z})=-f_{L}(e^{Z}e^{Z^{\dagger}}-1)[1+T_{m}f_{L}(e^{Z}e^{Z^{\dagger}}-1)]^{-1}e^{-Z^{\dagger}},\label{identity1}
\end{equation}
then multiplying the determinant by ${\rm Det}\,e^{Z^{\dagger}}=1$, and finally combining the product of three determinants into a single determinant. In this way we eliminate the matrix inversion, arriving at
\begin{align}
F(\xi)={}&\biggl\langle\prod_{m=1}^{N}
{\rm Det}\biggl(\bigl[1+T_{m}f_{R}(e^{-Z}e^{-Z^{\dagger}}-1)\bigr]e^{Z^{\dagger}}\bigl[1+T_{m}f_{L}(e^{Z}e^{Z^{\dagger}}-1)\bigr]\nonumber\\
&+T_{m}(1-T_{m})f_{R}(e^{-Z}-e^{Z^{\dagger}})f_{L}(e^{Z}e^{Z^{\dagger}}-1)\biggr)\biggr\rangle\nonumber\\
={}&\biggl\langle\prod_{m=1}^{N}{\rm Det}\biggl(1+T_{m}[(1-f_{R})e^{Z^{\dagger}}f_{L}-f_{R}e^{-Z}(1-f_{L})](e^{Z}-e^{-Z^{\dagger}})\biggr)\biggr\rangle.\label{Fxi3}
\end{align}
This is Eq.\ \eqref{Fxiresult} in the main text.
\end{widetext}

\newpage

\section{Derivation of the generating function in the shot noise regime}
\label{outline2}

Starting from the expression \eqref{Fxiresult2} for the generating function in the shot noise regime $k_{B}T\ll V$, we give the steps required to arrive at the bilinear form \eqref{Fxiresult3}. We group terms with $Z$ and with $Z^{\dagger}$ in the matrices $A_{m}=T_{m}f_{L}Z\bar{f}_{R}$ and $B_{m}=T_{m}f_{L}Z^{\dagger}\bar{f}_{R}-Z^{\dagger}$, so that Eq.\ \eqref{Fxiresult2} can be written as
\begin{equation}
F(\xi)=\bigl\langle\prod_{m=1}^{N}{\rm Det}(1+A_{m}+B_{m})\bigr\rangle.\label{Fxiresult2a}
\end{equation}
Because energies separated by $V^{p}$ with $p\geq 2$ can be discarded, we may set $A_{m}^{2}\rightarrow 0$, $B_{m}^{2}\rightarrow 0$. For any pair of matrices $A,B$ which square to zero, one has the identity
\begin{equation}
{\rm Det}\,(1+A+B)={\rm Det}(1-AB).\label{identity4}
\end{equation}
This leads to
\begin{align}
F(\xi)
={}&\biggl\langle\prod_{m=1}^{N}{\rm Det}\bigl(1+T_{m}Z\bar{f}_{R}Z^{\dagger}f_{L}\nonumber\\
&-T_{m}^{2}Z\bar{f}_{R}f_{L}Z^{\dagger}\bar{f}_{R}f_{L}\bigr)\biggr\rangle.\label{Fxiresult2b}
\end{align}

Eq.\ \eqref{Fxiresult3} follows by noting that $Z\bar{f}_{R}f_{L}\rightarrow Z\bar{f}_{R}$ for $k_{B}T\ll\Omega\simeq V$, since the Fermi function $f_{L}$ in this term is evaluated at energies near $E_{F}$, where it can be replaced by unity. Similarly $Z^{\dagger}\bar{f}_{R}f_{L}\rightarrow Z^{\dagger}f_{L}$, since $\bar{f}_{R}$ is evaluated at energies near $E_{F}+V$ where it can be replaced by unity.


\begin{thebibliography}{99}
\bibitem{Gla63} R. J. Glauber, Phys. Rev. \textbf{131}, 2766 (1963).
\bibitem{Man05} L. Mandel and E. Wolf, \textit{Optical Coherence and Quantum Optics} (Cambridge University, Cambridge, 1995).
\bibitem{Bee01} C. W. J. Beenakker and H. Schomerus, Phys. Rev. Lett. \textbf{86}, 700 (2001).
\bibitem{Gab04} J. Gabelli, L.-H. Reydellet, G. F\'{e}ve, J.-M. Berroir, B. Pla\c{c}ais, P. Roche, and D. C. Glattli, Phys. Rev. Lett. \textbf{93}, 056801 (2004).
\bibitem{Zak10} E. Zakka-Bajjani, J. Dufouleur, N. Coulombel, P. Roche, D. C. Glattli, and F. Portier, arxiv:1001.1411.
\bibitem{Bee04} C. W. J. Beenakker and H. Schomerus, Phys. Rev. Lett. \textbf{93}, 096801 (2004).
\bibitem{Leb09} A. V. Lebedev, G. B. Lesovik, and G. Blatter, arXiv:0911.4676.
\end{thebibliography}
\end{document}